\documentclass[aps,twocolumn]{revtex4}
\usepackage{amsmath}
\usepackage[ansinew]{inputenc}
\usepackage[english]{babel}
\usepackage{graphicx}
\usepackage{graphicx}
\usepackage{dcolumn}
\usepackage{amssymb}
\usepackage{amsmath}
\usepackage{epsfig}

\newcommand{\be}{\begin{equation}}
\newcommand{\ee}{\end{equation}}
\newcommand{\bea}{\begin{eqnarray}}
\newcommand{\eea}{\end{eqnarray}}

\def\lapp{\mathrel{\rlap{\raise.5ex\hbox{$<$}}
                    {\lower.5ex\hbox{$\sim$}}}}
\def\gapp{\mathrel{\rlap{\raise.5ex\hbox{$>$}}
                    {\lower.5ex\hbox{$\sim$}}}}

\def\M0{m_{0}}

\def\MSTAU1{m_{\tilde \tau_1}}
\def\MXI10{m_{\tilde \chi_1^0}}
\def\MST1{m_{\tilde t_1}}

\def\MHU{m_{H_u}^2}
\def\MHD{m_{H_d}^2}

\def\nn{\nonumber}

\def\sigu{\Sigma_u}
\def\sigd{\Sigma_d}

\def\sq2{\sqrt{2}}
\def\hlf{\frac{1}{2}}
\def\s2t{s_{2\theta}}
\def\C2t{c_{2\theta}}

\def\mz{m_{\scriptscriptstyle Z}}

\def\m0{m_0}

\def\gev{{\rm GeV}}

\newcommand{\gsim}{\lower.7ex\hbox{$\;\stackrel{\textstyle>}{\sim}\;$}}
\newcommand{\lsim}{\lower.7ex\hbox{$\;\stackrel{\textstyle<}{\sim}\;$}}

\begin{document}
\title {Are supersymmetric models with minimal particle content  under tension for testing at  LHC?}
\author
{\sf Abhijit Samanta\footnote{The correspondence email address: abhijit.samanta@gmail.com}}
\author
{\sf Sujoy Kumar Mandal}
\author
{\sf Himadri Manna}
\affiliation
{\em Nuclear and Particle Physics Research Centre, Department of Physics, Jadavpur University, Kolkata 700 032, India 
}
\date{\today}
\begin{abstract}
In supersymmetric models with minimal particle content and without large left-right squarks mixing, the conventional knowledge is that  the  Higgs Boson mass around 125 GeV leads to top squark masses ${\cal O}(10)$ TeV, far beyond the reach of colliders. Here, we pointed out that this conclusion is subject to several theoretical uncertainties. We find that electroweak symmetry breaking and  evaluation of Higgs mass at a scale far away from the true electroweak symmetry breaking scale introduce a large uncertainty in Higgs mass calculation. We show that the electroweak symmetry breaking at the scale near the true vacuum expectation value of Higgs field can increase the Higgs Boson mass about 4-5 GeV and can lower the bounds on squarks and slepton masses to 1 TeV.  Here we pointed out that the Higgs mass even with inclusion of radiative corrections can vary with electroweak symmetry breaking scale. We calculate it at two loop level and show that it varies substantially. We argue that Higgs mass  like other coupling parameters can vary with energy scale and the Higgs potential with all orders loop corrections is scale invariant. 
This  uncertainty  to the Higgs mass calculation due to electroweak symmetry breaking around 
the supersymmetry breaking scale, normally taken as $\sqrt{m_{\tilde t_L} m_{\tilde t_R}}$, to minimize the 1-loop radiative corrections can  be removed if one considers all significant radiative contributions to make Higgs potential renormalization group evolution scale invariant and evaluates electroweak symmetry breaking 
at the scale near the electroweak symmetry breaking scale. 
 A large parameter space becomes allowed when one considers electroweak  symmetry breaking at its true scale not only for producing correct values of the Higgs masses, but also for providing successful breaking of this symmetry in more parameter spaces.

\end{abstract}

\maketitle

\noindent

\section{\bf Introduction} 
The discovery of Higgs Boson  at ATLAS \cite{atlashiggs} and CMS \cite{cmshiggs} leads to  Higgs mass ($m_h$) calculation an important subject of impressive precision studies. In supersymmetric models with minimal particle content the tree level Higgs mass can not be larger than $m_Z \simeq$ 91 GeV. The large radiative corrections which are function of masses and couplings  of supersymmetric theories have direct implications on the discovery prospects of supersymmetry at colliders. But, there are many theoretical uncertainties in $m_h$ calculation
and  it needs to improve them for definite conclusion for the discovery prospect of supersymmetry at LHC. In this paper, we first pointed out that  electroweak symmetry breaking (EWSB) and  calculation of $m_h$ at the renormalization group evolution (RGE) scale far away from the  EWSB scale (which might be close to the vacuum expectation value (VEV) of Higgs field $(v_{\rm weak})$) introduce a large uncertainty in $m_h$ calculation.
 
We show significant increase in the mass of the CP-even neutral Higgs Boson $m_h$ if one evaluates  EWSB and calculates $m_h$ at  $Q_{\rm EW} \sim v_{\rm weak}$ instead of   $Q_{\rm EW} = \sqrt{m_{\tilde t_L} m_{\tilde t_R}}$. This leads to a dramatic change in the allowed parameter space in supersymmetric models.
 

The EWSB is considered at $Q_{\rm EW} = \sqrt{m_{\tilde t_L} m_{\tilde t_R}}$ in all  spectrum generator packages available in literature \cite{isajet, suseflav, softsusy, suspect, spheno} and also in finding post-LHC constraints \cite{msugra-lhc}.  This technique to evaluate EWSB at RGE scale other than the true EWSB scale is used to make radiative  corrections   negligible compared to the tree level Higgs potential. In these studies, $m_h$ is also calculated at $Q_{\rm EW} = \sqrt{m_{\tilde t_L} m_{\tilde t_R}}$. It has been shown in \cite{Dedes:2002dy} that the RGE scale dependence of the Higgs potential becomes negligible if one adds the  dominant  two loop corrections ${\cal O} (\alpha_t \alpha_s + \alpha_t^2)$ to the Higgs potential and it enables one to calculate EWSB and Higgs masses  at any scale other than the scale where the 1-loop corrections to the Higgs potential is negligible.

\section{\bf Radiative corrections to EWSB}
The tree level scalar potential keeping only the dependence on the neutral Higgs fields: 
\bea
V_0&&=\left(\MHU+\mu^2\right){\mid{H_u^0}\mid}^2+\left(\MHD+\mu^2\right){\mid{H_d^0}\mid}^2 +\nonumber\\ && {m_3}^2({H_u^0}{H_d^0}+h.c.)+\frac{g^2+g^{'2}}{8}({\mid{H_u^0}\mid}^2-{\mid{H_d^0}\mid}^2)^2. \eea  

Here, both the tree level potential $V_0$ and its parameters are strongly RGE scale dependent.
However,  in principle, if we include loop corrections at all orders,  the effective potential $V_{\rm eff} = V_0 + \Delta{V}$  should be RGE scale independent.
From the  minimization criteria one can find
\bea
\label{eqmu}
\mu^2 & = & - \frac{\mz^2}{2} 
+ \frac{m^2_{H_d} + \sigd - (m^2_{H_u} + \sigu) 
\,\tan^2\beta}{\tan^2\beta - 1}\,,\\
&\nn\\
\label{eqm3}
m_3^2 & = & -\hlf\,\sin 2\beta\, \left( m^2_{H_d} + m^2_{H_u} 
+ 2\,\mu^2 + \sigd + \sigu \right)
\eea
where,  
\be \Sigma_{a}= \frac{1}{2v_a} \frac{\partial \Delta V}{\partial v_a}.\ee
For simplicity, $\Delta V$ is not calculated separately, but one directly evaluates $ \Sigma_{a}$. The checking of convergence of the $V_{\rm eff}$ is not done through evaluation of $\Delta V$, but through the invariance of the value of $\mu$ with respect to EWSB scale $Q_{\rm EW}$. It ensures that perturbation series for  $V_{\rm eff}$ converges at all RGE scales.
The one loop corrections $\Delta{V_1}$ in Landau gauge 
is given by \cite{coleman-weinberg}:
\bea \Delta V_1 = \frac{1}{64\pi^2} STr M^4 \left [ \ln (M^2/Q^2) - 3/2\right ] \eea

The dominant contribution that comes from stop quarks is given by:
\bea \Sigma_u(\tilde t_i) \sim \frac{3 y_t^2}{16\pi^2 } m_{\tilde t_i}^2 ln (m_{\tilde t_i}^2/Q^2) \eea
The loop corrections are very significant, without which the evaluation of parameters from minimization of the tree level potential may give even wrong results \cite{Gamberini:1989jw}.
These radiative corrections depend strongly on the RGE scale $Q$  and the 1-loop contributions  normally become negligible at $Q = \sqrt{m_{\tilde t_L} m_{\tilde t_R}}$. 


The minima of effective Higgs potential  $V_{\rm eff}$ is strongly RGE scale dependent even with complete 1-loop corrections \cite{Martin:2002wn}.
But, the  addition of 2-loop corrections shows RGE scale invariance of the minima of the potential $V_{\rm eff}$ 
 and the  parameters obtained from this minimization criteria ($\mu$ and $m_3^2$) are  RGE scale invariant \cite{Martin:2002wn}. 
The RGE scale dependence of $V_{\rm eff}$ is shown in terms of $\mu$ (obtained from the  minimization of $V_{\rm eff}$) in Fig. \ref{f:q0-mu}
at tree level, 1-loop level and 2-loop level. It is seen that the scale dependence is almost completely    
 negligible at 2-loop level.

\begin{figure*}[htb]
\includegraphics[width=8.cm,angle=0]{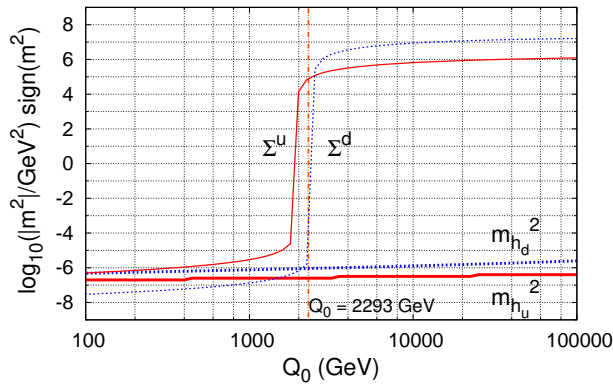}
\caption{The variations of $m^2_{h_{u,d}}$ and $\Sigma^{u,d}$ with electroweak symmetry breaking scale $Q_{\rm EW}$ for a typical  set of mSUGRA input parameters $m_0 = 600, m_{1/2} = 1500, A_0 = -1700, \tan\beta = 40, sign(\mu) = +1$. 
}
\label{f:sigud}
\end{figure*}

 Here, it should be noted that $\Sigma_{a}$ can be large at any scale and may  be even comparable with $m_{H_{u,d}}^2$ as it is the derivative of $\Delta V$ with respect to $v_a$.
The large  values of  derivative of $\Delta V$  do not mean the violation of convergence of  $V_{\rm eff}$. 
The Higgs mass squared parameters $\MHU$ and $\MHD$ are also not physically observable. The interaction of the   
the Higgs field with other fields  is such that these parameters $m_{H_{u,d}}^2, \Sigma_{u,d}$ can change rapidly with RGE scale by a few orders of magnitude from high positive to high negative value from GUT scale to weak scale (e.g., $+10^8$ GeV to $-10^8$ GeV  for a typical set of input parameters). This is shown in Fig. \ref{f:sigud}. 

%

%


\section{\bf EWSB scale and evaluation of Higgs mass}
The Standard Model
\cite{sm} identifies weak scale $Q_{\rm EW} = (\sqrt{2\sqrt{2}G_F})^{-1} = 175$ GeV with
the VEV of a fundamental, isodoublet, ``Higgs'' scalar field.
The minimization of the Higgs potential gives VEVs of the neutral part of the Higgs fields.    In MSSM, EWSB fixes  $\mu^2$ and $m_3^2$ as the VEV $v_{weak}=\sqrt{{\langle{H_u^0}\rangle}^2 + {\langle{H_d^0}\rangle}^2 }$ is already fixed from Standard Model predictions. 

The scale of supersymmetry breaking $Q_{\rm SB} \sim m_{\tilde S}$ (the masses of sparticles)  and the scale of EWSB     minima $Q_{\rm EW}$ originate from breaking of two completely separate symmetries  and the physics at these two scales are completely different.  

\begin{figure*}[htb]
\includegraphics[width=7.cm,angle=0]{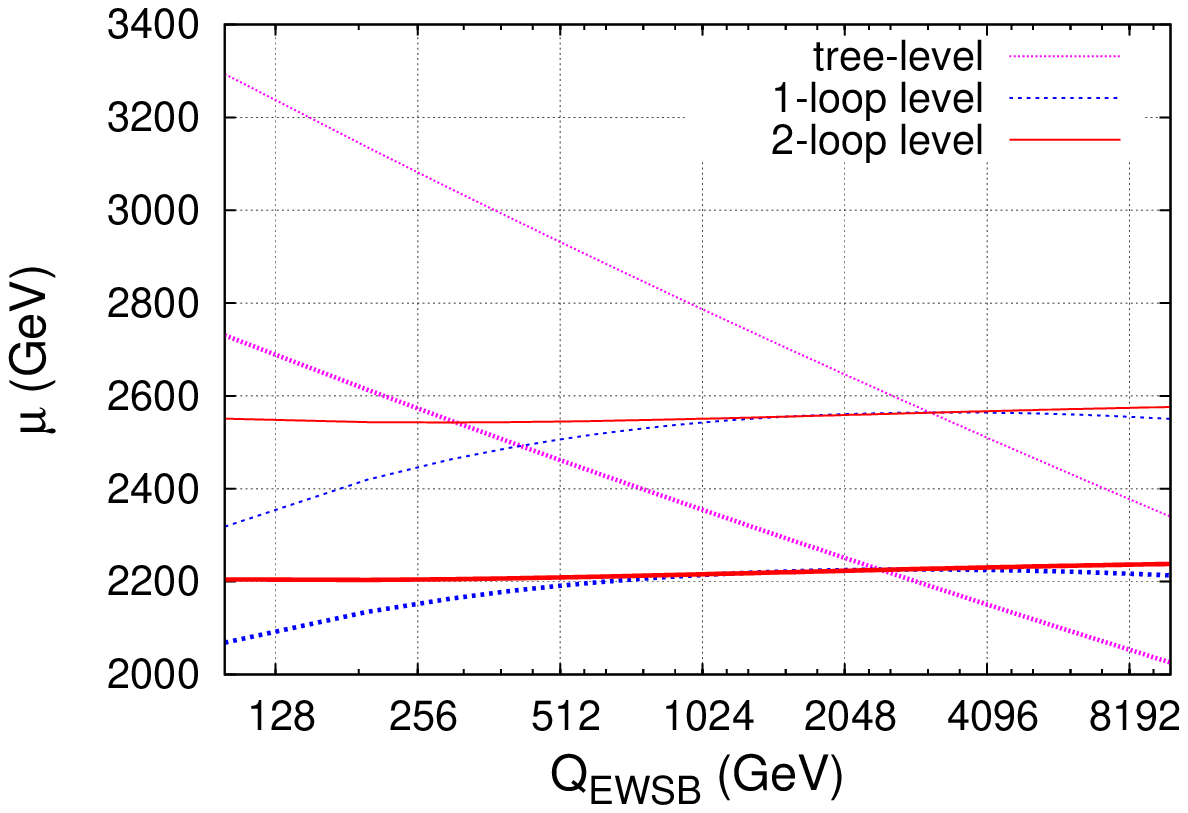}
\includegraphics[width=7.cm,angle=0]{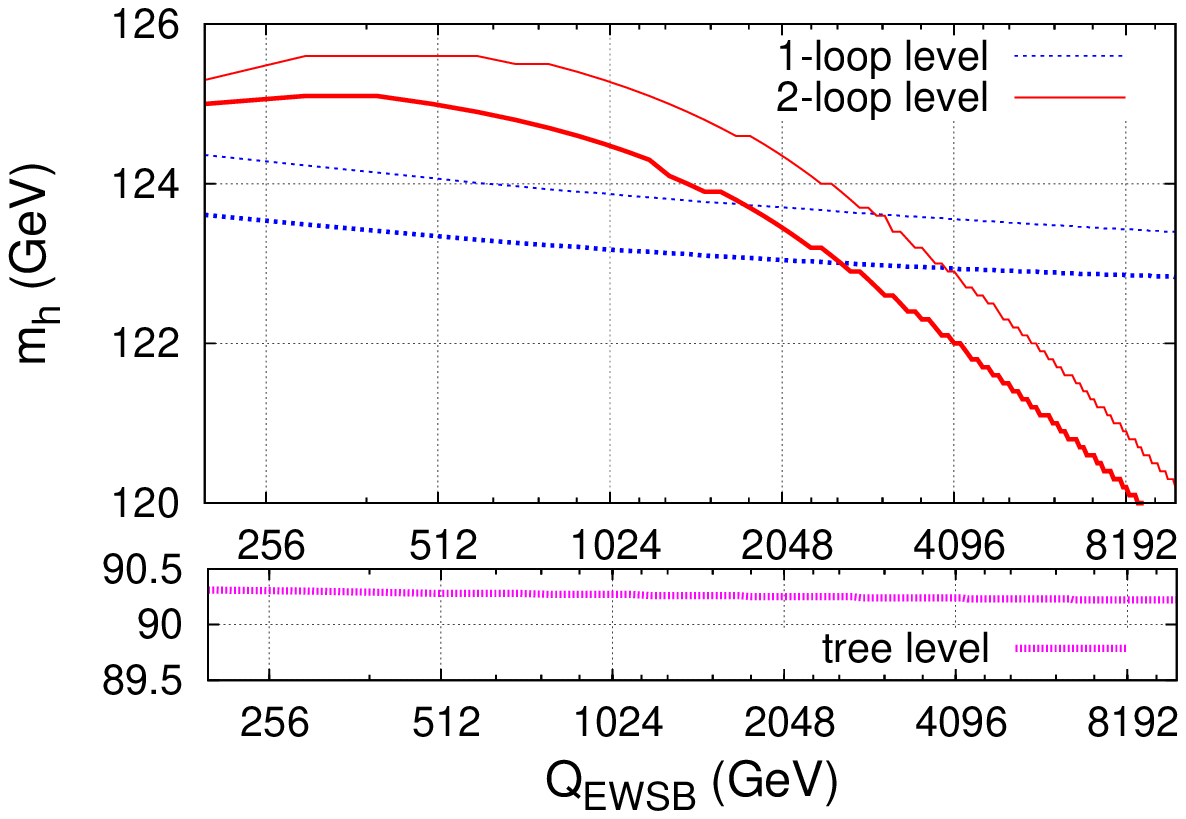}
\caption{The variation of $\mu$ (left) and $m_h$ (right) with the electroweak symmetry breaking scale $Q_{\rm EW}$ for  typical  sets of mSUGRA input parameters ($m_0 {(\rm GeV)}, m_{1/2}{(\rm GeV)}, A_0{(\rm GeV)}, \tan\beta, sign(\mu)) =(390, 2490, 0, 14.5, 1)$ (thinner lines),  and $(390, 1895, -1125, 14.5, 1)$ (thicker lines), respectively. The value of $\sqrt{m_{\tilde t_L} m_{\tilde t_R} }$ are 3817 GeV and 2900 GeV, respectively.   
}
\label{f:q0-mu}
\end{figure*}

 If  $m_{\tilde S} \sim$ a few hundred GeV, then the running of parameters up to 
$Q=v_{weak}$ is negligible and one can use $Q_{\rm SB}$ as the weak scale. But, if  $m_{\tilde S} \sim$ TeV, one cannot neglect the running of the parameters (particularly, $\MHU$ and $\MHD$)  and the approximation of using $Q_{\rm SB}$ as $Q_{\rm EW}$ does not work.  The value of $m_h$ is increased significantly when one evaluates EWSB at  $Q_{\rm EW}\approx  v_{weak}$ (see Fig. \ref{f:q0-mu}). On the other-hand, if one considers EWSB    scale $Q_{\rm EW} \sim$ TeV, EWSB  also  may not occur for some region of parameter space due to less running of $\MHD$ and $\MHU$ (EWSB requires  $\mu^2$  positive).


\begin{table*}
\begin{center}
\begin{tabular}{|c|c|c|c|c|c|c|c|}
\hline
{Input} &  &
\multicolumn{2}{|c|}{SuSeFLAV-1.2} &
\multicolumn{2}{|c|} {SuSpect 2}& 
\multicolumn{2}{|c|}{FeynHiggs-2.12.0}\\
&  &
\multicolumn{2}{|c|}{ $m_h$ (GeV) at}  &
\multicolumn{2}{|c|} {$m_h$ (GeV) at }& 
\multicolumn{2}{|c|}{$m_h$ (GeV) at }\\
$m_0 ({\rm GeV}), m_{1/2} ({\rm GeV}) , A_0 ({\rm GeV}), \tan\beta, {\rm sign}(\mu)$ &
 $\sqrt{m_{\tilde t_L} m_{\tilde t_R}}$ (GeV)&
$\sqrt{m_{\tilde t_L} m_{\tilde t_R}}$&
$v_{\rm weak}$& 
$\sqrt{m_{\tilde t_L} m_{\tilde t_R}}$&
$v_{\rm weak}$& 
$\sqrt{m_{\tilde t_L} m_{\tilde t_R}}$&
$v_{\rm weak}$ \\
\hline
$390, 2490, 0, 14.5, 1$ &
3817 & 123.0 & 125.3 & 123.8 & 126.9 &124.7&125.2\\
\hline
$390, 1895, -1125, 14.5, 1$ &
2900& 122.8 & 125.0 &123.5 &126.3&125.1&125.4\\
\hline
$513,     2321,    -1281,        6.2,       -1$ &
3507 & 120.1 & 124.9 &120.9&123.9&121.6&122.4\\
\hline
 $1956,      592,    -4128,       14.3,       -1$ &
992 & 123.0  & 125.0  & 123.7 & 125.4  & 122.7  & 123.7 \\
\hline
\end{tabular}
\end{center}
\caption{The typical values of $m_h$ calculated at $Q_{\rm EW} =  \sqrt{m_{\tilde t_L} m_{\tilde t_R} }$ and $Q_{\rm EW} \approx v_{\rm weak} = \sqrt{{\langle{H_u}\rangle}^2 + {\langle{H_d}\rangle}^2 }$ obtained from different program packages for different sets of inputs. The less differences in $m_h$ values between two scales appear for FeynHiggs since the RG running of Yukawa and other couplings are not considered here and they are same at these two scales. The input MSSM parameters are only different at two scales.}
\label{t1}
\end{table*}

In our calculation, we consider program SuSeFLAV-1.2 \cite{suseflav}. It considers full one loop corrections together with two loop leading contributions ${\cal O} (\alpha_t \alpha_s + \alpha_t^2)$ to the Higgs mass squared parameters following ref. \cite{Dedes:2002dy}. We have compared  SuSeFLAV-1.2 with softsusy3.4.0 \cite{softsusy} for different sets of input parameters and find no significant change; similar changes  in the spectra are observed with the changes in input parameters. 
For typical sets of mSUGRA \cite{msugra-model} input parameters we show in Fig. \ref{f:q0-mu} (left) that the variation of $\mu$ (evaluated at $Q_{EW}$) with $Q_{EW}$ at tree level, one loop level and two loop level. We find that $\mu$ is scale independent at two loop level. In Figure 2 (right) we show the variation of $m_h$ (evaluated at $Q_{EW}$) with $Q_{EW}$ at one loop level and two loop level. We find that $m_h$ remains scale dependent.

Here we pointed out that the Higgs mass (not potential) even with inclusion of radiative corrections can vary with electroweak symmetry breaking scale. We calculate it at two loop level and show that it varies substantially. We argue that Higgs mass  like other coupling parameters can vary with energy scale.

In Table \ref{t1} we compare the  values of $m_h$ calculated at the two scales $Q_{\rm EW} =  \sqrt{m_{\tilde t_L} m_{\tilde t_R} }$ and $Q_{\rm EW} \approx v_{\rm weak} = \sqrt{{\langle{H_u}\rangle}^2 + {\langle{H_d}\rangle}^2 }$ obtained from SuSeFLAV-1.2 \cite{suseflav},  SuSpect 2 \cite{Djouadi:2002ze} and FeynHiggs-2.12.0 \cite {Hollik:2014bua}. The less differences in $m_h$ values between two scales appear for FeynHiggs since the running of Yukawa and other couplings are not considered here and they are same at these two scales. The MSSM parameters are the input and they are considered to be different at these two scales. The EWSB at $Q_{\rm EW} = v_{\rm weak}$ is not only required for correct masses in the Higgs sector, but also for successful breaking of EW symmetry.  The parameter $\MHU$  goes to larger negative value as one decreases the RGE scale. This provides successful breaking of electroweak (EW) symmetry by yeilding $\mu^2$  positive. As a consequence a large parameter space becomes allowed. 

In brief, one can conclude that $V_{\rm eff}$ with complete radiative corrections  is RGE scale invariant and 
the accurate spectra through EWSB can  be found by generating them  at the true EWSB scale $Q_{\rm EW} \approx \sqrt{{\langle{H_u}\rangle}^2 + {\langle{H_d}\rangle}^2 }$. 
In generation of spectra through EWSB (masses of the Higgs particles), we run all MSSM parameters up to $Q_{\rm EW}$, and in generation of the masses of sparticles, all MSSM parameters are stored at $Q_{\rm SB}$. 


\section{\bf The mSUGRA parameter space} 
We have generated the allowed mSUGRA parameter space for  two  cases of evaluation of  EWSB     minima: i) at $Q_{\rm EW} =  \sqrt{m_{\tilde t_L} m_{\tilde t_R}}$ (scale considered for finding post-LHC constraint   in literature)  and ii) $Q_{\rm EW} \approx v_{weak} \approx \sqrt{{\langle{H_u}\rangle}^2 + {\langle{H_d}\rangle}^2 }$ (the true EWSB scale).
Here, we consider the only parameter space where one can generate $m_h = 125.5 \pm 0.5$ GeV. No other constraints  are considered (neutralino may not be the lightest supersymmetric particle (LSP)).
We generate the spectra for the range of $m_0 = 100 -3100 ~\gev$,  $m_{1/2} = 100 - 3100 ~\gev$ $A_0 = -3m_0~ {\rm to}~ + 3m_0$, $\tan\beta = 3 - 63$ and $sign(\mu) = \pm 1$. 

\begin{figure*}[htb]
\includegraphics[width=7.cm,angle=0]{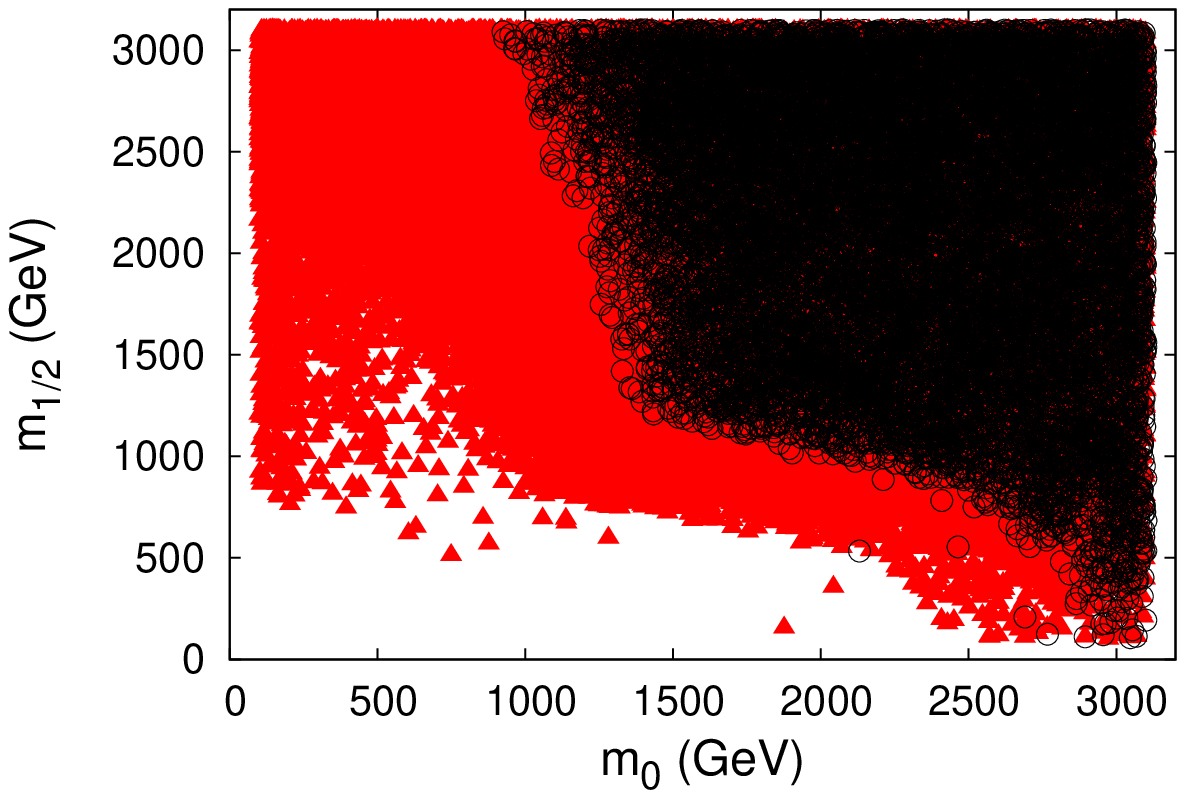}
\includegraphics[width=7.cm,angle=0]{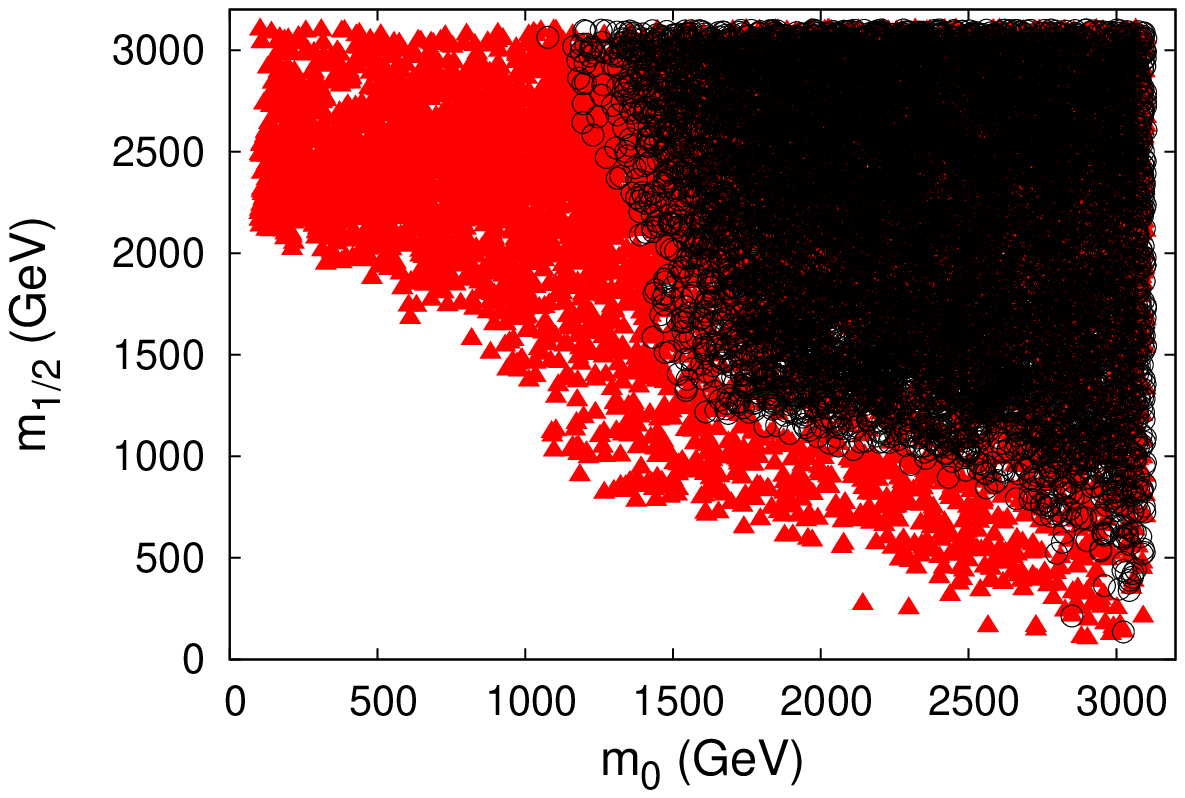}
\includegraphics[width=7.cm,angle=0]{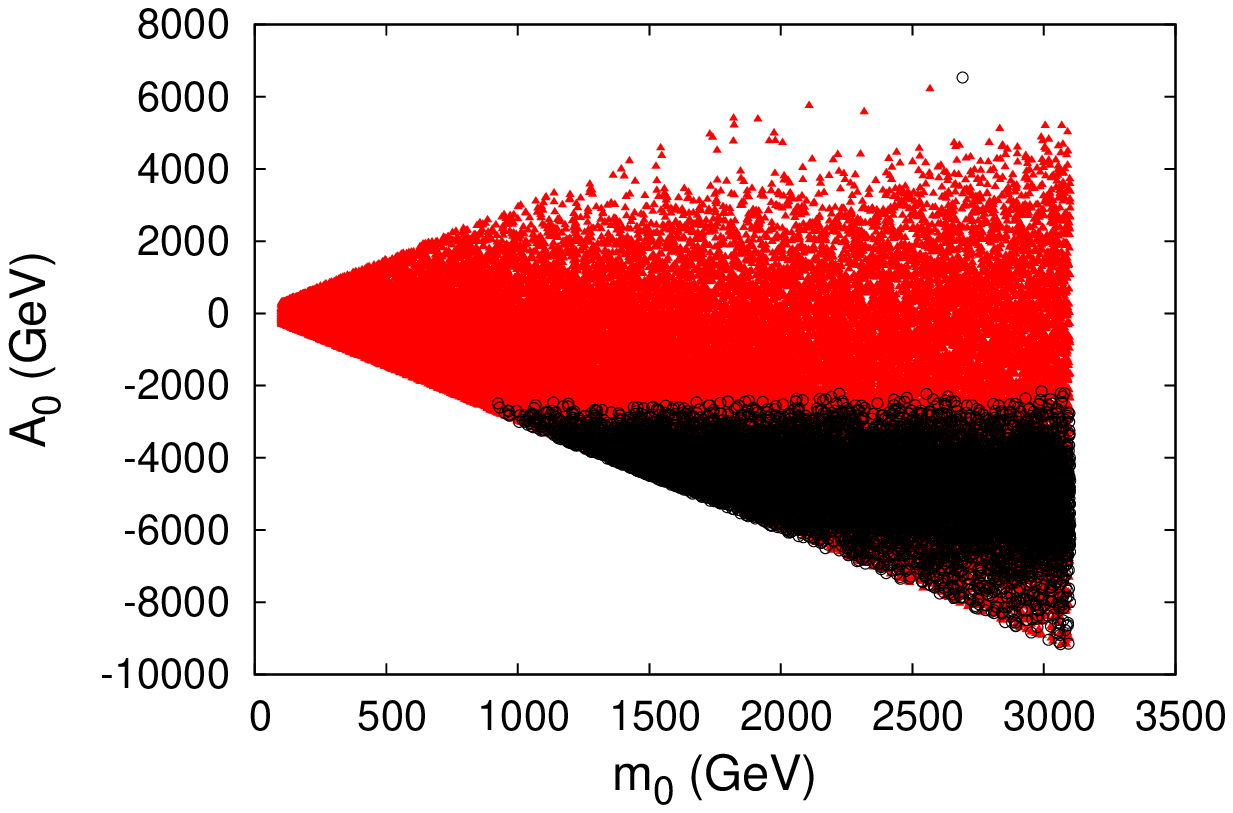}
\includegraphics[width=7.cm,angle=0]{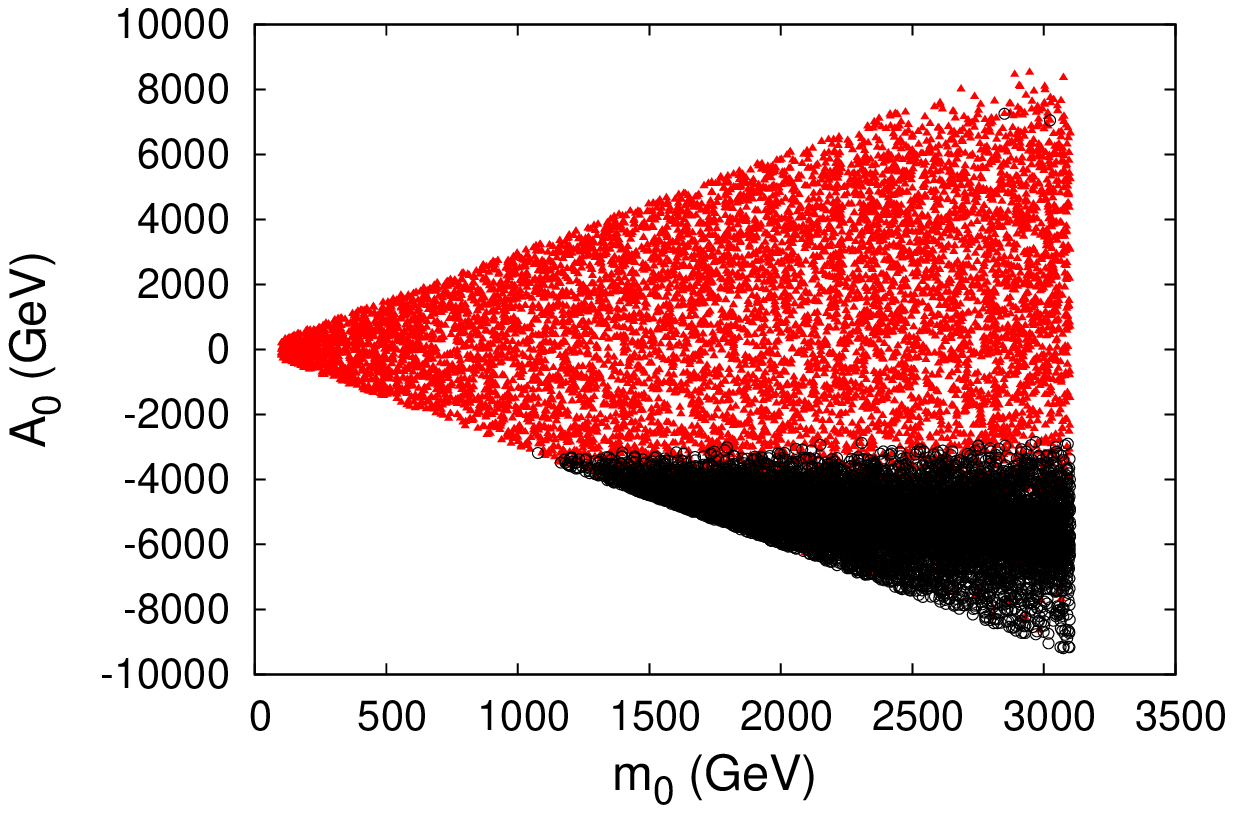}
\caption{\sf \small
The  allowed parameter space for $125 {\rm GeV} < m_h < 126$ GeV  in $m_0 - m_{1/2}$, $m_0 - A_0$ planes for sign($\mu$) positive (left) and negative (right), respectively. The point represented by triangle (circle) denotes EWSB at $Q_{\rm EW}=v_{\rm weak}$ ($Q_{\rm EW}=\sqrt{m_{\tilde t_L} m_{\tilde t_R}}$).
}
\label{f:m0-mhf}
\end{figure*}

In Fig. \ref{f:m0-mhf}, it is seen that a dramatically large parameter space in mSUGRA model  with almost no absolute bounds on $m_0, m_{1/2}$ and $A_0$ is allowed  when  EWSB  minima  is evaluated at  $Q_{\rm EW} \approx v_{weak}$  in contrary with the one when EWSB minima is evaluated at $Q_{\rm EW} =  \sqrt{m_{\tilde t_L} m_{\tilde t_R}}$.

The parameter $\MHU$  goes to larger negative value as one decreases the RGE scale. This also provides successful breaking of EW symmetry  in more parameter spaces producing $\mu^2$  positive and yeilding correct masses for Higgs particles. A large parameter space becomes allowed when one considers EWSB at its true scale.


%


\section{Conclusion}
In conclusion, we pointed out that   electroweak symmetry breaking and  calculation of $m_h$ at the scale other than the true vacuum expectation value of Higgs field introduces an uncertainty in Higgs mass calculation. One can remove this uncertainty if one considers all significant radiative contributions to make Higgs potential renormalization group evolution scale invariant and evaluates electroweak symmetry breaking 
at the true vacuum expectation value of Higgs field.
Then, there will be  no strong {\it absolute}  bounds on $m_0$, $m_{1/2}$ and $A_0$ in mSUGRA model to produce $m_h$ around 125 GeV. We pointed out that the Higgs mass (not Higgs potential) even with inclusion of radiative corrections can vary with electroweak symmetry breaking scale. We calculate it at two loop level and show that it varies subtantially. Finally, we argue that Higgs mass  like other coupling parameters can vary with energy scale. A large parameter space becomes allowed when one considers EWSB at its true scale not only for producing correct value of the Higgs mass, but also for providing successful breaking of EW symmetry in more parameter spaces.

{\bf Acknowledgments:} The author AS is grateful to Scientific and Engineering Research Board, Department of Science and Technology, Govt. of India for opening the scope of doing research through financial support under the research grant SB/S2/HEP-003/2013.

\bibliography{bibtex/nu}

\end{document}